\documentclass[aps,prc,twocolumn,showpacs,superscriptaddress,floatfix,10pt]{revtex4}
\usepackage{bm}
\usepackage{amssymb}
\usepackage{amsmath}
\usepackage{graphicx}
\begin{document}

\title{Dark matter transport properties and rapidly rotating neutron stars}

\author{C.J. Horowitz}
\email{horowit@indiana.edu}
\affiliation{
Department of Physics and CEEM, Indiana University, Bloomington, Indiana 47405, USA
}%
\date{\today}

\begin{abstract}
Neutron stars are attractive places to look for dark matter because their high densities allow repeated interactions.   Weakly interacting massive particles (WIMPs) may scatter efficiently in the core or in the crust of a neutron star.  In this paper we focus on WIMP contributions to transport properties, such as shear viscosity or thermal conductivity, because these can be greatly enhanced by long mean free paths.   We speculate that WIMPs increase the shear viscosity of neutron star matter and help stabilize r-mode oscillations.  These are collective oscillations where the restoring force is the Coriolis force.  At present r-modes are thought to be unstable in many observed rapidly rotating stars.  If WIMPs stabilize the r-modes, this would allow neutron stars to spin rapidly.  This likely requires WIMP-nucleon cross sections near present experimental limits and an appropriate density of WIMPs in neutron stars.

\end{abstract}

\pacs{
95.35.+d, 
97.60.Jd, 
26.60.Dd, 
66.20.-d  
}

\bigskip

\maketitle
Dark matter has proven to be elusive.  Weakly interacting massive particles (WIMPs) have likely not yet been seen in the laboratory.  In astrophysics, neutron stars may be promising places to look for WIMPs because they are the only known objects dense enough for WIMPs to interact repeatedly.    WIMPs could condense to form a black hole and destroy their host neutron star \cite{kouvaris,kouvaris2, kouvaris3,lavallaz,guver}.  Perez-Garcia et al. suggested that WIMPs could convert a (hadronic) neutron star into a strange star \cite{angeles}.  Alternatively, WIMP annihilation could heat neutron stars \cite{kouvaris4,kouvaris5,lavallaz}.  Finally Leung et al. considered the impact of dark matter on the frequencies of neutron star oscillation modes \cite{leung,leung2}.  

In this paper we focus on WIMP contributions to transport properties.  A few WIMPs may greatly increase the shear viscosity or thermal conductivity of neutron star matter because WIMPs have very long mean free paths $\lambda_W$ that efficiently transport momentum and energy.  For example $\lambda_W$ may be $10^{10}$ times longer than the mean free path of an electron.  This means that, per particle, a WIMP will contribute dramatically more to the shear viscosity.  Note that by focussing on transport properties we are taking advantage of the long WIMP mean free path.  This is the very thing that often makes studying dark matter so difficult.     

We start with a model where $\lambda_W$ is smaller than the size of a neutron star.  Next we calculate the WIMP contribution to the shear viscosity of dense neutron rich matter and speculate that this shear viscosity may damp r-mode oscillations in rapidly rotating neutron stars.  Finally we discuss WIMP contributions to the thermal conductivity.

WIMPs could be trapped in the core or in the crust of a neutron star.  We consider both possibilities in turn.  First, let us assume $\lambda_W$ is dominated by scattering from nucleons in the core of a neutron star
\begin{equation}
\lambda_W = \frac{1}{\sigma_{Wn} n_n},
\label{mfp}
\end{equation}
with $\sigma_{Wn}$ an effective transport cross section for WIMP-nucleon scattering that includes the effects of Pauli blocking, see below.  The nucleon density is $n_n$.  We assume $n_n\gg n_W$ where $n_W$ is the WIMP density and we neglect WIMP self-interactions.  We require that  $\lambda_W$ is not much larger than the size of the star.  Consider 
\begin{equation}
\lambda_W\approx 10\ {\rm km}.
\label{mfp10km}
\end{equation}
For simplicity we model a neutron star as a uniform density sphere with say $n_n \approx 2n_0=0.32$ fm$^{-3}$.  Here $n_0$ is normal nuclear density.  Note, the central density of a neutron star is expected to be significantly larger than $n_0$.  Equation \ref{mfp10km} requires 
\begin{equation}
\sigma_{Wn}\approx 3\times 10^{-45}\ {\rm cm}^2\, .
\label{sigma}
\end{equation}
For $M_W$ near 50 GeV and assuming spin independent interactions, this is slightly below -- but near -- the recent experimental limits of the CDMS and EDELWEISS \cite{CDMS} and XENON 100 \cite{Xenon100} experiments.  For larger or smaller masses, Eq. \ref{sigma} is significantly below the experimental limits.  

Nucleon Pauli blocking may reduce the effective cross section $\sigma_{Wn}$ in the star so that $\lambda_W$ is too large.  In this case spin dependent interactions could still produce the required mean free path, since they are much more weakly constrained by experiment.  Alternatively, WIMPS could be trapped in the inner crust of the neutron star because of coherent WIMP-nucleus scattering.  Although the density is lower in the crust, the coherent cross sections may be large.

We now consider WIMP scattering in the neutron star inner crust.  Coherent WIMP nucleus scattering cross sections are proportional to $A^2$ where $A$ is the mass number.  We choose to normalize this cross section per nucleon.  Therefore coherence increases the cross section per nucleon by a factor of $A$.  This coherence is reduced by the form factor of the nucleus $F(q)$ with $q$ the momentum transfer and $F(q=0)=1$, $F(q\rightarrow\infty)=0$.  In addition, in the dense medium other nuclei can screen the scattering.  This is described by the static structure factor $S_A(q)$.  The static structure factor adds coherently scattering from all of the different nuclei \cite{S(q)}.  Screening ensures that $S_A(q=0)$ is small and the static structure factor is normalized $S_A(q\rightarrow \infty)=1$.

The differential WIMP-nucleus scattering cross section (per nucleon) in the medium $d\sigma/d\Omega$ is
\begin{equation}
d\sigma/d\Omega = d\sigma/d\Omega|_0 A\, S_A(q)\, F(q)\, .
\label{sigma_medium}
\end{equation}
Here $d\sigma/d\Omega|_0$ is the free space WIMP-nucleon scattering cross section.  Integrating over scattering angles gives the total cross section
\begin{equation}
\sigma=\int d\Omega\, d\sigma/d\Omega = \sigma_0 \langle A \rangle
\label{sigma_tot}
\end{equation}
Here $\sigma_0=\int d\Omega\, d\sigma/d\Omega|_0$ and Eq. \ref{sigma_tot} defines the effective number of nucleons 
\begin{equation}
\langle A \rangle = A\, \langle S_A(q) F(q) \rangle \,
\label{Aave}
\end{equation}
In general $\langle A \rangle < A$ because $S_A(q)<1$ for small $q$ and $F(q)\le 1$.  We expect the product $S_A(q)F(q)$ to peak at intermediate momentum transfers near $q^*\approx 0.3$ fm$^{-1}$ \cite{pasta}.  If the momentum transfer range for the integral in Eq. \ref{sigma_tot} includes $q^*$ then $\langle A \rangle$ may only be somewhat smaller than $A$ say $\approx A/2$ \cite{pasta}.  However if the WIMP momentum is too small, so that $q^*$ is not part of the integration range, than $\langle A \rangle$ can be $\ll A$ because $S_A(q)\ll 1$ for small $q$.

We now estimate the optical depth $\tau$ for WIMP scattering in the inner crust.
\begin{equation}
\tau = \int dr \sigma_0 \langle A \rangle n_n(r) \approx \Delta r\, \sigma_0\, \langle n \rangle\, \langle \bar A \rangle 
\label{tau}
\end{equation}   
Here $\Delta r\approx 500$ m is the thickness of the inner crust, the nucleon density is $n_n(r)$ and $\langle n(r) \rangle \approx 0.05$ fm$^{-3}$ is a typical density in the inner crust.  Finally $\langle \bar A \rangle$ is $\langle A \rangle$ appropriately averaged over the composition in the inner crust.  If the WIMP momentum $p$ is large enough so that $2p>q^*$ then a very rough estimate is $\langle \bar A\rangle\approx 100$.  The mass number $A$ of nuclei present in the inner crust tends to increase with density.  In addition, $A$ depends on the assumed nuclear interaction.  For example Thomas Fermi calculations of Douchin et al. \cite{haenseleos} or relativistic mean field calculations of G. Shen et al. \cite{sheneos} tend to have larger $A$ than semiclassical simulations \cite{pasta0}.  We will discuss these estimates further in a later paper, see also \cite{pasta}.    

With these crude estimates the optical depth of the inner crust is $\tau \approx \sigma_0 \, 2.5 \times 10^{18}\ {\rm fm}^{-2}$.  Therefore an optical depth of order one requires a cross section 
\begin{equation}
\sigma_0 \approx 4 \times 10^{-45}\ {\rm cm}^2\, .
\label{sigma_0}
\end{equation}
Note this is close to Eq. \ref{sigma}.  For simplicity we assume $M_W$ is (much) less than the mass of the nucleus.  In that case, a WIMP of momentum $p$ striking a nucleus at rest has a maximum momentum transfer of $2p$.  For WIMPs in thermal equilibrium at temperature $T$ one has $p\approx (3kTM_W)^{1/2}$.  We require $2p\ge q^*$.  For $M_W=50$ GeV this is $T\ge 7\times 10^7$ K, or $3.5\times 10^8$ K for $M_W=10$ GeV.  At lower temperatures than these, WIMPs may travel too slowly to scatter efficiently in the crust.  We conclude that it is possible for dark matter to have a significant optical depth in the inner crust.  However this likely requires WIMP cross sections near present experimental bounds.

WIMPs may be captured, thermalized, and trapped using a combination of scattering in the core and in the crust.   When a WIMP first nears a neutron star, gravity accelerates it to a high energy.  These energetic WIMPs can scatter efficiently in the core because nucleon Pauli blocking plays less of a role.  The energetic WIMP can easily excite nucleons above the Fermi sea so $\sigma_{Wn}\approx \sigma_0$.  However, energetic WIMPs can scatter at large $q$  where the nuclear form factor $F(q)\ll 1$.  Therefore scattering in the crust may be inefficient for high energy WIMPs.  Over time the WIMPs can loose energy by repeatedly scattering from nucleons in the core until the WIMPs have much lower energy.  At that point Pauli blocking is important so scattering in the core becomes inefficient $\sigma_{Wn}\ll \sigma_0$.  However low energy WIMPs can now scatter efficiently in the crust.

We turn now to transport properties and the contribution of WIMPs to the shear viscosity of neutron star matter.  For simplicity we consider just the core of a neutron star.  We model the core as composed of dense neutron rich matter and a low density gas of WIMPs.  From simple kinetic theory, the shear viscosity of this WIMP gas $\eta_W$ is
\begin{equation}
\eta_W = \frac{1}{3}(3kTM_W)^{1/2} \lambda_W n_W\, .
\label{etaW}
\end{equation}    
Here $n_W$ is the number density of WIMPs.  

If the mean free path in Eq. \ref{mfp10km} is achieved, let us compare the shear viscosity in Eq. \ref{etaW} to the shear viscosity of neutron rich matter $\eta_n$ without any WIMPs.  At a density of $2n_0=0.32$ fm$^{-3}$ and a temperature $T\approx 10^8$ K (which is typical for a LMXB) Shternin et al. calculate $\eta_n=2\times 10^{19}$ g cm$^{-1}$ s$^{-1}$ \cite{shternin}.  This is the sum of electron (dominant), muon, and nucleon contributions in a non-superfluid star.  For $M_W=50$ GeV, $\eta_W$ in Eq. \ref{etaW} will be equal to $\eta_n$ if the density of WIMPs is 
\begin{equation}
n_W\approx 3\times 10^{28}\ {\rm cm}^{-3}\, .
\label{rhoW}
\end{equation}
This corresponds to a number fraction of WIMPs to nucleons $X_W$ of
\begin{equation}
X_W=\frac{n_W}{n_n}\approx 10^{-10}\, .
\label{X_W}
\end{equation}
{\it We conclude that WIMPS will likely dominate the shear viscosity of dense neutron rich matter if their number fraction $X_W$ is larger than this value.}   This estimate is for the core of the star.  We expect qualitatively similar estimates for the inner crust.  However this should be verified in future work.  The density of WIMPs in the core of a neutron star may depend on the age of the star, the distribution of dark matter in the galaxy, and many WIMP properties such as mass, boson/fermion nature, and interactions with both normal matter and other WIMPs.  If WIMPS significantly enhance the shear viscosity this will increase the damping of collective oscillations such as r-modes and may impact the maximum spin rate of neutron stars.  

How can neutron stars spin so fast?  For example, a pulsar in Terzan 5 spins at 716 Hz \cite{716Hz}.  
This is puzzling because rapidly rotating neutron stars appear to be unstable to r-mode oscillations \cite{rmode_puzzle}.  These modes are a class of oscillations where the restoring force is the Coriolis force \cite{rmode_review}.  The emission of gravitational waves can excite r-modes and cause the amplitude of oscillations to grow.  Bildsten and Andersson et al. first suggested that r-modes could provide a limit on the spin of neutron stars in low mass X-ray binaries (LMXB) \cite{bildsten}.  Here angular momentum gained from accretion is radiated away in gravitational waves.  This keeps the stars from spinning faster.  Indeed, gravitational waves from r-modes may be detectable in large scale interferometers such as LIGO \cite{ligo}.

An r-mode is unstable if the rate energy is gained through gravitational wave radiation (tapping the stars rotational energy) exceeds the rate energy is lost through damping due to shear and bulk viscosities.  This helps motivate many calculations of the shear viscosity of neutron matter \cite{benhar}, neutron rich matter with electrons and muons \cite{shternin, issac},  nuclear pasta with complex non-spherical shapes \cite{pasta}, and a crust-core boundary layer \cite{boundary,boundary2}. In addition, the bulk viscosity of neutron rich matter \cite{bulkneutron,bulkneutron2, issac}, hyperon matter \cite{hyperons, hyperons100,hyperons101,hyperons102,hyperons103,hyperons104,hyperons105}, and quark matter \cite{q106,q107,q108,q109,q110,q111,q112,q113,q114,q115,q116} has been calculated.  The damping from all of these different sources appears to be too small to stabilize the r-modes in many observed neutron stars \cite{rmode_puzzle}.  {\it This problem is solved if WIMPs increase the shear viscosity enough to stabilize the r-modes.}




The viscosity in Eq. \ref{etaW} depends only weakly on $M_W$ and $T$.  However, heavy WIMPs may travel a long time between collisions and have a long relaxation time.  This could imply a frequency dependence for $\eta_W$ if the WIMPs can not keep up with an r-mode, whose frequency is of order the rotation frequency of the star.  This frequency dependence of $\eta_W$ may depend more strongly on $M_W$ and $T$ and could also lead to a nonzero bulk viscosity.   Possible WIMP contributions to the bulk viscosity should be examined in future work.  WIMPs can also contribute to the thermal conductivity.  This may not be important for the core of the star because the thermal conductivity is thought to be very high even with out WIMPs.  However, WIMPs could have an impact on crust cooling after extended periods of accretion \cite{crustcooling,crustcooling1,crustcooling2,crustcooling3,crustcooling4}.  Here accretion heats the crust and then the crust is observed to cool rapidly after accretion stops.  This rapid cooling suggests a very high crust thermal conductivity that is likely consistent with (or without) an additional WIMP contribiton.  In addition WIMPs could transport heat in the crust of a strongly magnetized star where it may be otherwise difficult to transport heat perpendicular to the magnetic field.  This could lead to more equal surface temperatures and less time dependence to the thermal X-ray flux as the star rotates. 

We speculate that observations of neutron star rotation may provide indirect information on dark matter.  For example young neutron stars may have captured little dark matter and as a result their r-modes may become unstable at relatively low frequencies. Therefore one may not expect to observe young rapidly rotating stars.  In contrast old neutron stars may have captured more dark matter,  stabilizing their r-modes.  Indeed a number of old neutron stars are observed to spin quickly as millisecond pulsars.    Alternatively, neutron stars may spin quickly in regions of the galaxy with a high dark matter density, such as the galactic center.  Note that the neutron star with the fastest known rotation is near the galactic center \cite{716Hz}.
  
In summary, neutron stars are attractive places to look for dark matter because their high densities allow repeated interactions.   Weakly interacting massive particles (WIMPs) may scatter efficiently in the core or in the crust of a neutron star.  In this paper we focused on WIMP contributions to transport properties because these can be greatly enhanced by long WIMP mean free paths.   We speculate that WIMPs increase the shear viscosity of neutron star matter and help stabilize r-mode oscillations.  As a result, neutron stars can spin rapidly.  This likely requires WIMP-nucleon cross sections near present experimental limits and an appropriate density of WIMPs in neutron stars.

We thank Ed Brown, Jinfeng Liao, and Gerardo Ortiz for helpful comments.  This research was supported in part by DOE grant DE-FG02-87ER40365.

\end{document}